# The coupling of thermodynamics with the organizational water-protein intra-dynamics driven by the H-bonds dissipative potential of cluster water

Alfred Bennun[1, 2]

[1]Full Professor - Emeritus - Rutgers University, USA
[2]CONICET, AR

## ABSTRACT

Physiological integration allows the Red cell-Hb-CSF to function as a sensor adapting response to Hb heterotropic equilibriums. At the lungs the mutual inclusion of $O_2$ and $Mg^{2+}$, each one increasing affinity for the other, stabilizes the relax (R) form in a $[(O_2)_4Hb(Mg)_2].(H_2O)_R$ complex. At tissue level, the inclusion of $H^+$ and 2,3-DPG couples for exclusion of $O_2$ and $Mg^{2+}$ to stabilize the tense (T) form in a 2,3-DPG-deoxyHb-$(H_2O)_T$ complex. The system T into R and R into T integrates both senses into a cycle, without involving a direct reversal like T<-->R. Open thermodynamics allows the dissipative potential of water cluster $(H_2O)n$ to interact with the hydrophilic asymmetries of Hb, to restrict the kinetic sense randomness of a single peak for activation energy ($Ea$). Instead the conformational dynamics of hydration shells could sequence an enhanced $Ea$ into several peaks, to sequentially activate intermediate transitions states. Hence, $\Delta\mu$ (dipole states), $\Delta$sliding, $\Delta$pKa, $\Delta$n-H-bonds, etc., could became concatenated for vectoriality. $(H_2O)n$ by the loss of H-bonds couple with to the hydration turnover of proteins and ions to result in incomplete water cluster $(H_2O)n^*$, with a lower "n". $(H_2O)n^*$ became a carrier of heat/entropy into the cerebrospinal fluid (CSF) which has to be replaced 3.7 times per day. OxyHb formation involves sliding of $\beta_1\alpha_1$ vs $\beta_2\alpha_2$, to shift $\alpha_1$ and $\alpha_2$ Pro 44 into allowing the entrance of a fully hydrated $[Mg.(H_2O)_6](H_2O)_{12-14}^{2+}$ (or $Zn^{2+}$) into the hydrophilic $\beta_2\alpha_1$ and $\beta_1\alpha_2$ interfaces. OxyHb pKa of 6.4 leads to $H^+$-dissociation increasing negative charge of R-groups. This at $\beta_2\alpha_1$ sequence two tetradentate chelates, first a $Mg^{2+}$, enters into coordinative bonding with $\beta_2$ His 92 and a second $Mg^{2+}$ with $\alpha_1$ His 87, to cooperatively release hindrance. The interconversion of oxy-to-deoxyHb, pKa=8, leads to the amphoteric imidazole to became positively charged and proximal histidines return into hindrance position, releasing incompletely hydrated $[Mg(H_2O)inc]^{2+}$ and $O_2$ into CSF. The protonated $\beta_1$ and $\beta_2$ His 143 are released from the chelates to salt-link with 2,3-DPG.

## INTRODUCTION

Photophosphorylation studies did not allow detection of reversibility in the absence of un-couplers [1] [2], indicating the presence of a restriction to microscopic reversibility. Reconstitution of the architecture of lipid-protein regions within a membrane was shown to be required for the binding of CF1-ATPase [3].

The dynamics of H-bonds on the hydration shells of ATPase were shown capable to modify the catalytic activity of the enzyme [4]. Energy transduction was shown to be mediated by high energy conformational intermediate coupling across coordinative bonding [5]. These were characterized as delocalizing activation energy ($Ea$) for a vectorial sequencing of transition states [1] [2] [3].

Prigogine (1947) [6] proposed that a system with a minimum entropy production, could steadily go down to a minimum were it stays, supporting for a time a steady state. This non-linear probable state refers to perturbations of thermodynamic forces, too small to invalidate a principle of maximum entropy production [7].

A system is closed when a boundary, allows passage of work and heat. The system is open, when matter and energy can pass across the boundary. Phase boundaries could characterize an organismal relationship with its environment. The uptake of metabolite is mediated by an air phase boundary, allowing uptake of $O_2$ and release of $CO_2$ and Heat. The latter, became a carrier of entropy to the outside of the organismal system.

Large hadron colliders have produced a rather large number of high energy particles, which could be plotted within a range from $10^{-25}$ to $10^{-7}$ seconds, according to their half-life decay, which became an enthalpy contribution to the arrow of time [8]. These dissipative events generate neutrinos and antineutrinos which scape the open system by lack of return reactivity.

In linear systems entropy production is the possibility of cosmic evolution to support self-organization in other regimes of local dynamics [9] [10] [11] [12].

Microscopic reversibility of dynamics implies that the matrix for events is symmetric [13] and generates a single *Ea* peak for either the forward or reverse sense of the reactions.

Prigogine's premise: "Dynamics and thermodynamics limit each other" [14] indicates that these processes could be differentiated. The forms of the enzyme within membrane allows transition complexes which would be maintained, under asymmetric phase angles, between conformational structure dynamics for coupling and thermodynamic equilibrium and kinetics [15].

The organization of molecules within defined membrane structures imposes physicochemical constraints, which are not especially subject to a random distribution of *Ea* [16]. By the contrary in a lipid-enzyme membrane the inter- and intra-molecular kinetic energy could be represented by coupling between several peaks driving specific transitions states, along the vectorial progress of an energy transduction process [16].

Intra-molecular asymmetry could be confers by water dynamics capacity to differentiate a hydrophilic space, or region from a hydrophobic one [4]. This allows that the hydration shell of proteins and ions could confer turnover to water architectures when couple with the dissipative potentials of water cluster $(H_2O)n$ [4]. Hence, the H-bonds formed by a $-\Delta G$ of restructuring hydration shells enhances the energy requirements to reach transition state.

The four Heme groups of Hb show a 2-fold symmetric axis [17] [18] [19] leading to the idea that any one Heme site, would interact equally over the 24Å inter Heme to Heme distances for cooperative $O_2$-ligation [20].

The two states concerted MWC model (Monod-Wyman-Changeux model) [21] proposed an allosteric mechanism which in stereochemical basis was assumed to implicate that the relative ratio of the equilibrium between low vs. high $O_2$ affinity forms of Hb, could represent conformational forms, a tense (T) versus a relax (R) [22]. Protein dynamics explain the allosteric behaviors of hemoglobin. The study of deoxyHb revealed a presence of a central cavity for 2,3 DPG binding [23] [24] and the tendency of oxygenation to induces the association of $Mg^{2+}$ [25] [26] or $Zn^{2+}$ [27] to Hb.

A homotropic system (crystals) was used by Perutz to show that the breaking of salt-links could trigger a one-way T to R change of Hb. The implicated on the alkaline Bohr's effect were the $\beta_1$ and $\beta_2$ His 146 becoming protonated to salt-link $\beta_1$ and $\beta_2$ Asp 94 [22].

The binding of O2 (as well as allosteric effectors, such as protons and organic phosphates) altered the relative stabilities of the T (low $O_2$ affinity) and R (high $O_2$ affinity). The hindrance by the salt-bridges in the T form lead proximal histidines to restrain the movement of the iron atom within the porphyrin plane required for oxygen binding to Hemes. At the beta-Hemes, the distal valine and histidine block the oxygen-combining site in the T-structure. Oxygenations rupture of salt-bridges in T allowing a pKa decrease in oxyHb [28].

R-residues of Hb participate to form by mutual exclusion, either a chelating site or a binding site for 2,3-DPG, increasing and decreasing affinity for $O_2$, respectively [29].

The amphoteric response of histidines in deoxy to oxy allows the imidazole side chains to acquire a negative charge N and by $C\alpha$-rotation to attract divalent metals to form chelating site at the $\beta_2\alpha_1$ and the $\beta_1\alpha_2$ interface [29] [30]. Pulling out both β His 143 from conforming the central cavity releases 2,3 DPG. Thus, allowing to postulate that $Mg^{2+}$ and $Zn^{2+}$ compete with 2,3 DPG in a mutually exclusive manner [29][30] [31].

## RESULTS

### 1. Structure and Function

However, Hb structure shows hydrophobic regions at $\beta_1\alpha_1$ and $\beta_2\alpha_2$ intradimer interfaces and hydrophilic polar R-groups at the $\beta_2\alpha_1$ and $\beta_1\alpha_2$ interfaces [31] [32].

During oxygenation the hydrophilic asymmetry of Hb allows a fully hydrated $Mg^{2+}/Zn^{2+}$ ($[Mg(H_2O)_6](H_2O)_{12}$) to enter first into the $\beta_2\alpha_1$ interface for specific sequential chelation of R-groups. The process leads the hydrated metal to an exergonic binding with Hb compensated by the endergonic loss of most of the divalent metal hydration shell [33].

Human-RBC hemolysates show an increment of $^{14}CO_2$ released from the consumption of $[1-^{14}C]$-glucose, within the hexose monophosphate (HMP) pathway by the $Mn^{2+}$ stimulation of the redox recycling of NADPH (nicotine adenine dinucleotide phosphate), which contri-bute to homeostasis of pH, because otherwise could favor lactic acid formation [34].

The level of glucose at the red cell could act as an integrated sensor of the brain needs, because in the Rapapport-Luebering scheme the 2,3-DPG phosphormutase, which is inhibited by low pH, becomes maximally activated at pH=7.4. Since cerebrospinal fluid (CSF) is slightly alkaline could signal the red cell to increase the erythrocyte level of 2,3-DPG [35] to form 2,3-DPG-deoxyHb-$(H_2O)_T$ (Figure 1) and release of $O_2$ to match glucose uptake and maintain aerobic glycolysis in brain generating the ATP required to operate $Na^+$-pump [36]. OxyHb in transition to deoxy releases a $[Mg(H_2O)inc]^{2+}$ in CSF, which by binding proteins to the membrane could protect the tendency of $ATP^{4-}$ to subtract $Mg^{2+}$ from the protein lipid structure [3] of neuron.

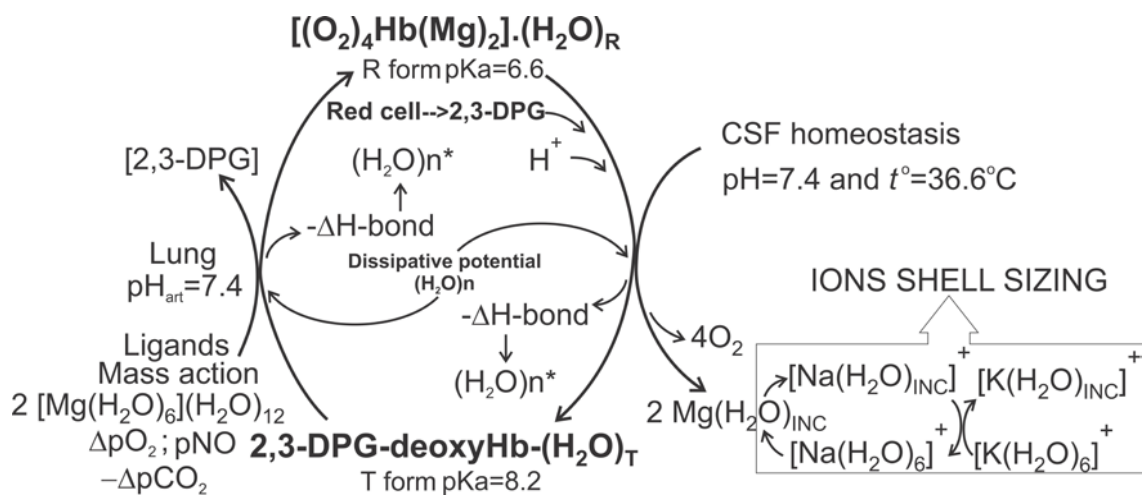

**Figure 1: Configuration of turnover from oxy- into deoxy- and return to oxy-Hb.** An open system supports in steady state the reactivity of water by its H-bond dissipative potential from incoming cluster-rich $(H_2O)n$, n=12-14, versus its exit as water cluster-poor $(H_2O)n^*$, n=5-6. The sub-indexing is use to indicate that the number of water molecules in their hydration shells allows to differentiate the "R" for relax and the "T" for tense forms of Hb. Interconversion of the deoxy- to oxy- depends on sliding shifting α Pro 44 allowing entrance at the two hydrophilic interfaces βα of $2[Mg(H_2O)_6](H_2O)_{12}$. The $Mg^{2+}$ ion losses most of hydration shell when binding Hb and during deoxygenation is released with an incomplete hydrated shell $[Mg(H_2O)inc]^{2+}$. Hence, its smaller size allow to be discharged at CSF and interact for ion shell sizing decreasing the hydration shell of $Na^+$: $[Na(H_2O)inc]^+$ which enters into its channel at the $Na^+$-pump to take $H_2O$-out of the hydration shell of $K^+$: $[K(H_2O)_6]^+$.

Figure 1 shows that the dissipative potential of $(H_2O)n$ can be maintained in steady state when an open system allows its input in the reaction media to be balanced by the output of exhausted water cluster $(H2O)n^*$ [36] [37] [38] [39].

During deoxygenation $H^+$-uptake became associated to the amphoteric response of protonating the imidazole of His residues. The $NH^+$ in the rings attracts $H_2O$ molecules for an exergonic H-bonding within deoxyHb, balancing the break (endergonic) of the coordinative bonds between $Mg^{2+}$ and Hb. Thus, releasing from chelating state, the $[Mg(H2O)inc]^{2+}$ with a high spontaneous tendency to subtract $H_2O$ from the hydration shells of protein an ions, allowing sizing for the fitting of ions into their respective gates. This may provide the $-\Delta G$ input required to build hydration shells even if CSF maintains a thermal-homeostatic [40].

CSF is produced in clusters at the thin walled capillaries called chorideplexes that line the walls of the ventricles. Its high water turnover maintains allostasis of cerebral water with cluster sizes of about 12 molecules [41].

Temperature is a colligative measurement and the endergonic process of H-bond breaking in $(H_2O)n$ is incomplete in the $(H_2O)n^*$ state. The reduction in the contained number of water in the cluster by decreasing its size may ease cell hydration. The increase in the internal vibrational state of molecules in an incomplete H-bonded network, could contribute within CSF, to maintain its homeostatic temperature.

Figure 1 shows that the turnover process breaks the electrical bonding of $H_2O$ molecules in $(H_2O)n$ from n=12-14 to about n=5-6 [41] [42]. Thus, $(H_2O)n^*$ by decreasing the number of H-bonds within the cluster traps a heat-attenuated carrier of the increment in entropy, to be released-out of the reaction boundary. However, outside the body in contact with lower temperature, the dipole tendency of $H_2O$ [43] slowly but spontaneously will allows reconstitution of $(H_2O)n$ state.

The steady state of available cluster rich water demands a 3.7 times of the 160ml volume of CSF/24hs about 600ml of CSF, matching the brain requirement of 20% of total body metabolic activity.

Figure 1 the interconversion of deoxy- to oxy- involves Hb conformational dynamics which progress driven by coupling to a sequence of thermodynamics events. Binding of divalent metal and $O_2$ and releasing $H^+$ and 2,3-DPG, in which $Ea_1$ drives Hb coordinative chelation and loss of water from the hydration shell of the divalent metal [33]. Thus, resulting in active transitions states in which the changes on hydration shell "architecture" of proteins and ions, ($\Delta$H-bonds), dipolar states ($\Delta\mu$), $H^+$ association ($\Delta$pKa) are chained by a delocalizable enhanced $Ea$ and dominance of each transition over a restricted span of time.

Therefore, delocalization of $Ea$ allows to overcome microscopic reversibility because implies a time asymmetric kinetic vector for water dynamics in the hydrophobic and hydrophilic differentiable inner space of the Hb molecule.

## 2. Characterization of R-groups reactivity tendency in Hb

Sliding of the dimer $\beta_1\alpha_1$ versus $\beta_2\alpha_2$ displaces $\alpha$ Pro 44 $\alpha_1$ increases the size of the central cavity allowing binding of 2,3-DPG, a characteristic of the T form. In deoxyHb the area of contact between $\beta_2$ and $\alpha_1$ chain shows $\beta_2$ His 97 rest in between the hydrophobic residue $\alpha_1$ Thr 41 and the zwitterionic $\alpha_1$ Pro 44 [20] blocking access to the hydrophilic crevice in the interface $\beta_2\alpha_1$ (Figure 2.A) of a fully hydrated $[Mg.(H_2O)_6](H_2O)_{12-14}^{2+}$, by symmetry a similar relationships is maintained for $\beta_1\alpha_2$. In deoxyHb the Bohr protons are more strongly associated in $\beta_1/\beta_2$ Cys 93 and Asp 94 [44] with less tendency to complex with $Mg^{2+}$ or $Zn^{2+}$ [20].

Proline is prevented from turning around its C$\alpha$ because the single N in the cyclic structure, is bound to two alkyl groups in $-60°$ dihedral angle $\varphi$ (phi, involving the backbone atoms C'-N-C$\alpha$-C'), which unstrained allows conformational rigidity [45]. The transition states of proline within peptide bonds $\varphi=\pm90°$ requires that the partial double bond be broken with $E_a$ of 20 kcal/mol. At body temperature amide groups can isomerize about the C-N bond between the cis and trans forms in about 3 seconds. $Mg^{2+}$ and Hb are competitors binding reciprocally with organic phosphates in the red cell. Deoxygenation increases the concentration of free $Mg^{2+}$ ion [46] which is released as the smaller incompletely hydrated specie $[Mg(H_2O)inc]^{2+}$, which can cross the $\alpha_1$ and $\alpha_2$ Pro 44 blockage in deoxyHb (Figure 1 and 2.A).

Rotation and sliding during oxygenation pulled the C-termini of $\beta$ chains away from contact with $\alpha$ chains and $\beta$ His 97 shifted in between $\alpha$ Thr 41 and $\alpha$ Thr 38 [20] (Figure 2.B). Hence, in oxyHb down-sliding unlocks $\alpha$ Pro 44 shifting it into to a position in which cannot longer sieve the access of fully hydrated $[Mg.(H_2O)_6](H_2O)_{12-14}^{2+}$ into the hydrophilic interface. $Mg^{2+}$ is required to activate adenylyl cyclase (AC) [47] [48], its product cAMP [49] and/or cGMP [50] could be transported into human erythrocyte (Human-RBC), against significantly large differentials of the intracellular-extracellular concentrations [51] [52], and therefore could be involved in feedback modulation.

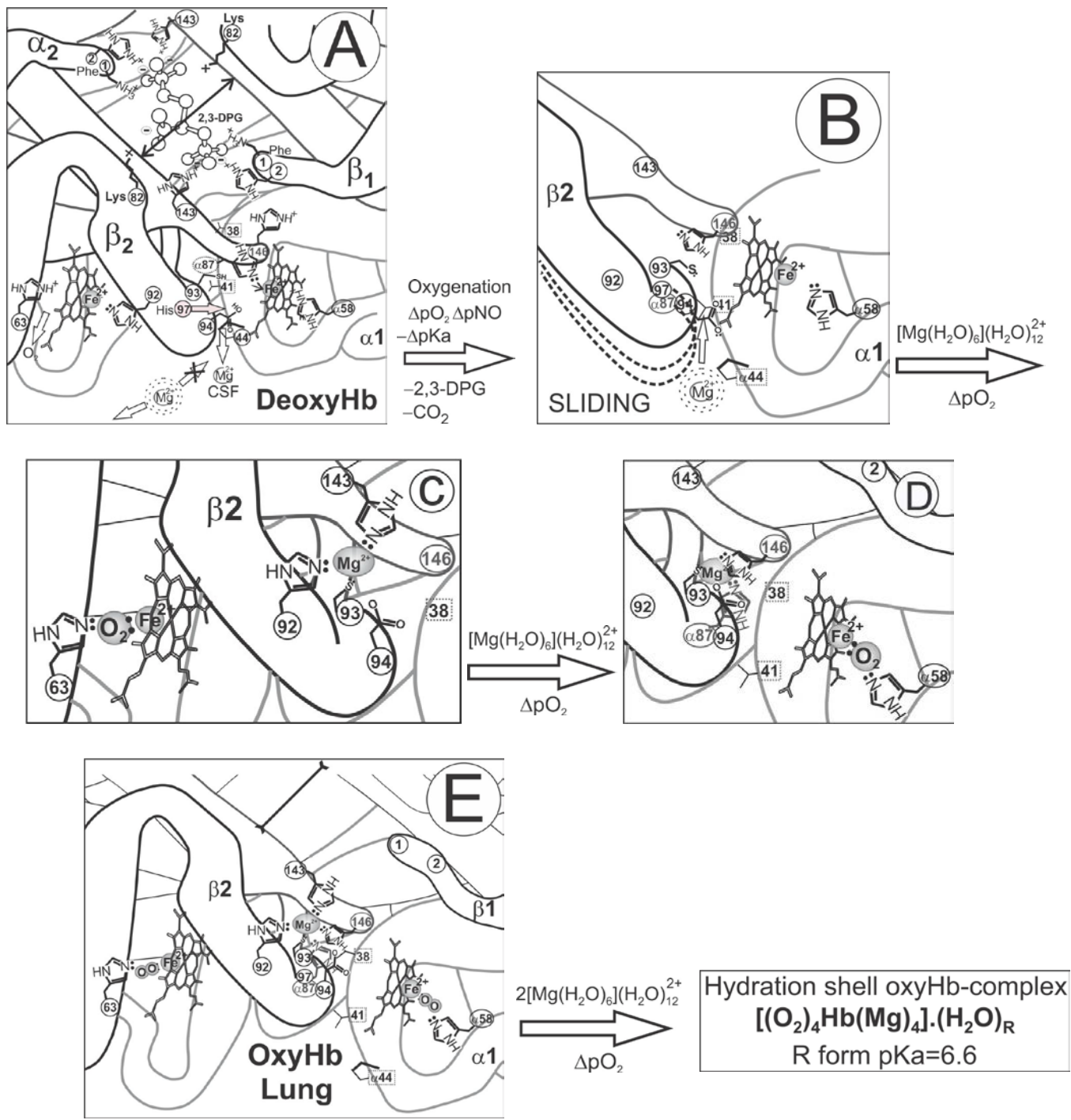

**Figure 2. Model sequencing reactivity changes during transition from deoxy- to oxyHb. A) 2,3-DPG-deoxyHb-$(H_2O)_T$.** The sieving effect: $\alpha_1$ Pro 44 when in between $\beta_2$ His 97 and $\alpha_1$ Thr 41 blocks the entrance of the larger fully hydrated $Mg^{2+}$ into the hydrophilic interface of $\beta_2$ and $\alpha_1$ chains, but allows the exit of the smaller incompletely hydrated $[Mg(H_2O)inc]^{2+}$. **B) Oxygenation,** at the lungs a pH=7.4, $\alpha$-chain sliding vs $\beta$-chain and entrance of fully hydrated $Mg^{2+}$ and/or $Zn^{2+}$ into hydrophilic region $\beta_2\alpha_1$ competitively shrinks the central cavity releasing 2,3-DPG. **C) Chelation:** ligand $Mg^{2+}$ increasing affinity of $\beta_2$ Heme for $O_2$. **D) Cooperativity:** The allosteric mechanism of a second $Mg^{2+}$ incrementing affinity of $\alpha_1$ Heme for $O_2$. **E) $[(O_2)_4Hb(Mg)_2].(H_2O)_R$** Events at dimer $\beta_2\alpha_1$ are subsequently and symmetrically reproduced at the dimer $\beta_1\alpha_2$.

Thus, helping to dissolve and eliminate waste and toxins. Integrated within an open system, water enters with a larger cluster size than the one eliminated.

The intradimer interfaces at $\alpha_1\beta_1$ and $\alpha_2\beta_2$ are hydrophobic. The figure 2 describes only the hydrophilic interface $\beta_2\alpha_1$ and do not show its symmetric interface $\beta_1\alpha_2$. Both contain Polar Regions and water with tendency to partially dissociate but initially $\beta_2\alpha_1$ is the more water reactive [53].

Figure 2.B) shows the $\alpha$-chain down-sliding vs the $\beta$-chain. The $\beta_2$ His 97 tends to became adjacent in the position between Thr 38 and Thr 41 of $\alpha_1$ chain, but $\alpha_1$ Pro 44 shift out of contact with $\beta$ chain [20]. The fully hydrated $[Mg(H_2O)_6](H_2O)_{12}$ can move inside the hydrophilic region [54].

In deoxyHb the imidazole ring of histidines could bear two NH bonds equal distributed by resonance structures in between both N. Hence, at the central cavity the negative charged 2,3-DPG binds by interaction with positive charged $\beta_2$ and $\beta_1$ His 143 [20].

OxyHb pKa=6 releases $H^+$ from histidines and breaks the salt-link between $\beta_2$ His 146 with $\beta_2$ Asp 94 [20]. The negative charged residues of $\beta_2$ Cys 93 and $\beta_2$ Asp 94 are attractants to the fully hydrated $[Mg(H_2O)_6](H_2O)_{12}$ [20].

In figure 2.C) it is shown that the divalent metals $Mg^{2+}$ and $Zn^{2+}$ are selected over monovalent ions as ligands of the negative atoms of hydrophilic R-groups within Hb. Multi-coordination forming chelate structures are manifested by many proteins. Hence, it is predicted that C$\alpha$-Rotation induce conformational changes which allow to progress from coordinative binding of bi-, to tetra- and even hexa-dentate chelation of $Mg^{2+}$. To reach this state the divalent metal has to loss most of their hydration shell. Therefore when deoxygenation release this ion would be only partially hydrated $[Mg(H_2O)inc]^{2+}$. Hence, Hb is proposed to fulfill the role not only of $O_2$ carrier but also of $Mg^{2+}$. This dehydrated cation could be a specific $H_2O$ acceptor to down size the hydration shells of other ions and proteins (Figure 1).

An initial $Mg^{2+}$ bidentate by coordination with $\beta_2$ Cys 93 and $\beta_2$ Asp 94 could be expanded to tetradentate by C$\alpha$-rotation of $\beta_2$ His 92 shifting away from its hindrance position, to allow $O_2$ to become a ligand at the $\beta_2$ Heme. Also $\beta_2$ His 143 by shifting from its position could initiating shrinkage of the central cavity and release of 2,3-DPG [20].

The covering of the allosteric distance between each $\beta$ and $\alpha$ Hemes requires hexadentate coordination by one $Mg^{2+}$ at the $\beta_2\alpha_1$ interface and a 2nd one at the $\beta_1\alpha_2$ interface. Figure 2.D) the stoichiometry of 4 divalent metals ($Mg^{2+}$ or $Zn^{2+}$) as ligands per Hb molecule could predict a sequence attracting 2 fully hydrated $[Mg(H_2O)_6](H_2O)_{12}$ first at $\beta_2\alpha_1$ and thereafter another two fully hydrated ions at the $\beta_1\alpha_2$ interfaces, which in the process reduce their hydration shells [33].

The 2nd divalent metals could incorporate by C$\alpha$-rotation the ligands $\beta_2$ Cys 93, $\beta_2$ Asp 94 plus $\beta_2$ His 146 for subsequently attract into the tetradentate coordination proximal $\alpha_1$ His 87, allosterically releasing hindrance to allow a second $O_2$ to become a ligand at the $\alpha_1$ Heme.

The coupling of a H-bond increment at the hydration shell of proteins and ions decreases entropy in the system itself, but increases it at the level of $(H_2O)n$.

The electrostatic attractions between molecules of water forming H-bonds have a tendency to be spontaneous, because involve the release of heat. The latter, is a slow process because considerable re-accommodation is required for the molecules of water to bond each other [43]. The latter may occur out of the system boundary, when wasted water enters in contact with the cooler air from outside the body system. Laboratory workers are usually aware that recently distilled water, has capture heat and allow several hours before its use, in order to prevent unreliable results.

## DISCUSSION

Closed enzyme-systems thermodynamics, at a body constant temperature, allows reactions to function according to *Ea*, to overcome the barrier of the differences in energy of formation between substrate (S) and product (P), and that of its coupled systems.

As catalysts enzymes (E) do not participate in the reaction stoichiometry. The protein structure itself is not consumed. However, it could be assumed that hydration shells change first in the direction of $E.(H_2O)n$ binding S to form the $E-S.(H_2O)n\pm xH_2O$ complex and after the reaction forming $E-P.(H_2O)n\pm yH_2O$ and when releasing P to complete turnover to free $E.(H_2O)n$. The hydration shell of some enzymes may determine turn-on versus -off states. The hydration states of the enzyme within the E-S vs E-P complexes requires H-bonds structural turnover in equilibrium with the dynamics of $(H_2O)n$. Hence, enzyme-protein conformational dynamics could uptake H-bonds consuming the dissipative potential between cluster states of surrounding water.

ATPase contributes to confer free energy to the overall thermodynamics of energy transduction process [15] [16]. Water dynamics may be complementary involved, since has been shown that ATPase activity is dependent on the hydration state of the protein [4].

The release of the divalent metal from Hb during deoxygenation is endergonic process requiring the amphoteric response turning the imidazole N atoms of R-His groups from negative into to positive. If these are not involve in forming salt-links the positive charges tend to form H-bonds, with the water taken-out from $(H_2O)n$.

Hence, coupling to water dynamics drives an overall exergonic process for Hb releasing $[Mg(H_2O)inc]^{2+}$ into CSF.

The dynamics of the hydration shells structure of ions and proteins could be correlated as structures that through dissipative process, generate enthalpy and lead reactions to completion, but turnover of the hydration shell requires water dynamics supported by a decrease in "n" of $(H_2O)n$ [4] [37] [55].

An $[Mg(H_2O)inc]^{2+}$ [33] could initiate H-bond capture from fully hydrated $Na^+$ leaving an incomplete hydrated shell around $Na^+$. The chaotropic $[Na(H_2O)inc]^+$ uptakes water from the fully hydrated $K^+$. The process decreases the effective ion size of $Na^+$ and $K^+$ concatenating sieving effects for ion translocation at the $Na^+/K^+$-pump [34] (Figure 1).

Additionally free $Mg^{2+}$ is required to activate basal AC and for norepinephrine binding to its receptor in AC, to bind to its receptor hormones like oxytocin and for attachment into the neuronal membrane of proteins generated by the cAMP Response Element-Binding (CREB) transcription [36].

Water mobility of the bulk-$(H_2O)n$ increases by the transfer of H-bonds decreasing mobility in the hydration shell of Hb [37]. Water dynamic implies acquisition of a latent activator potential, which allows the energy of breaking and reconstitution of H-bonds, to become a source for enthalpy release during formation of the transition states of Hb [36] [38] [39].

At room temperature thermal processes affect R-groups translational, vibrational, and rotational kinetics at the level of 0.8kcal/mol. The breakdown of H-bonds in $(H_2O)n$ could be evaluated for the bond O-H…:O as 5kcal/mol and for HO-H…:$OH_3^+$ as 4kcal/mol. The decrease in "n" within $(H_2O)n$ could couple with the increase (exergonic) in the number of H-bonds within a protein or an incomplete hydrated ion. These events increase tendency to drive the progress of transition states with a vector-sense. Therefore, coupling $Ea$ with water dynamics provide an energy level, which can overcome thermic randomness. Moreover, allows the increment in the H-bonds within a molecule to retain a coupling potential for a longer time that allowed by the heat dissipation of $Ea$.

## CONCLUSIONS

A homotropic system (crystals) was used by Perutz to show that the breaking of salt-links could trigger a one-way T to R change of Hb [18] [19] [20] [22]. The tendency to reach equilibrium by mass-action of reactants characterizes close systems, in which a single peak for $Ea$, allows microscopic reversibility. However, an open system is required to support a steady state conformational turnover of the protein.

Hence, the need to discern for vectorial systems a mechanism, which could prevents microscopic reversibility from entangling the directionality of processes lead to analyze Hb in function of multiple-equilibrium including $(H_2O)n$. Open systems couple the continuous entrance of S and its P exit, with a dissipative potential $-\Delta G$ to operate at far from the system final thermodynamic equilibrium. A thermodynamic dissipative potential of $O_2$ and metabolites uptake with $CO_2$ releases out of the system for entropy exclusion.

The overcoming of microscopic reversi-bility requires the integration of thermal and water dynamics, acting over the asymmetric reactive ten-dencies of Hb, under low vs. high $pO_2$. In a asymmetrically structured system an energy dissipative potential spread by coupling, between several transitions states could restrict reversibility. Maxwell description of an entropy reducing process required the separation of molecules by their temperature. In deoxyHb the sliding of $\beta_1\alpha_1$ vs $\beta_2\alpha_2$ allows $\alpha_1$ Pro 44 into closer contact with the $\beta_2$-chain and $\alpha_2$ Pro 44 with $\beta_1$-chain. Hecne, blocking the entrance of $[Mg(H_2O)_6](H_2O)_{12}$, but allows the exclusion of $[Mg(H_2O)inc]^{2+}$ and $(H_2O)n*$. Oxygenation shifts both α Pro 44 from contact un-blocking access to both hydrophilic regions. Hb sieving mechanism allows separating molecules by H-bonds (energy) and sizes.

After $Ea$ of sliding dissipates the molecule of Hb could no longer acquire the transition states capable to allow microscopic reversibility, because the R-groups which were previously in a reactive proximity shift into a more distant no longer reactive position. The separated of unidirectionality formed 2,3-DPG-deoxyHb-$(H_2O)_T$ to $[(O_2)_4Hb(Mg)_2].(H_2O)_R$ complex and viceverse functions in both senses as a exergonic ligand-sequence by coupling to H-bonds dissipative potential of $(H_2O)n$.

Not only the mass-action of ligand at the lungs vs CSF separate oxygenation from deoxygenation but also the dissipative clustering potential of $(H_2O)n$ into $(H_2O)n*$, which allows structural dynamics, with the capability to transfer free energy for the turnover of hydrated molecular architectures, well surpassing a simplistic radiator role.

Hence, CSF became replaced at a rate of three and a half times its volume per 24 hours helping to dissolve and eliminate toxins and $(H_2O)n*$. $H_2O$ enters in CSF with a larger cluster size than the one eliminated, to function as a carrier of entropy [56] to the outside of the open system reactive phase boundaries.

Rutgers University Connection Card/ID Office Information

Dr. Alfred Bennun                 Emeritus
647 East 14th Street, Apartment 2G     Expiration: Indefinite
New York, NY 10009-3158
ALFR9@HOTMAIL.COM
http://www.ncbi.nlm.nih.gov/pubmed?term=Bennun%20A.
http://www.biomedexperts.com